\documentstyle[mprocl]{article}

\def\be{\begin{equation}}
\def\ee{\end{equation}}
\def\bea{\begin{eqnarray}}
\def\eea{\end{eqnarray}}

\begin{document}

\title{Fractals and Symbolic Dynamics as Invariant Descriptors of
Chaos in General Relativity}

\author{ N.J. CORNISH }

\address{DAMTP, Cambridge University, Silver Street, Cambridge, CB3
9EW, UK}


\maketitle\abstracts{The study of dynamics in general relativity has
been hampered by a lack of coordinate independent measures of chaos.
Here I review a variety of invariant measures for quantifying
chaotic dynamics in relativity that exploit the coordinate
independence of fractal dimensions and symbolic entropies.
}
  
\section{Time and chaos}
Historically, chaos theory was developed for Newtonian dynamics where
time and space are absolute and the notion of a mechanical phase space
is clear. In contrast, both space and time are dynamical and
intermixed in general relativity. There is no such thing as {\em the}
time direction. The fundamentally different
role played by time in relativity and Newtonian mechanics manifests
itself in the coordinate, or gauge, dependence of chaotic
measures such as Lyapunov exponents\cite{os}. Lyapunov exponents
quantify a system's sensitive dependence on initial conditions. If
two initially close trajectories separate along a given eigendirection
in phase space such that the separation $\varepsilon(t)$ grows as
$
\varepsilon(t)=\varepsilon_{0} e^{\lambda t} \, ,
$
then $\lambda$ represents the Lyapunov exponent along that direction.
If $\lambda>0$ for a set of trajectories with non-zero measure,
the system is said to exhibit sensitive dependence on
initial conditions with a characteristic chaotic, or Lyapunov, timescale
$T_{L}=1/\lambda$. Unfortunately, this nice picture breaks down when
applied to general relativity. Consider the allowed coordinate transformation
$t \rightarrow \ln \tau $. In terms of this time variable we find
$
\varepsilon(\tau)=\varepsilon_{0} \tau^{\lambda} \; ,
$
which describes the standard power-law divergence of trajectories found in
integrable system. In particular, the Lyapunov exponents in
this coordinate system would all be zero. It should be mentioned
that the Lyapunov exponents also depend on the choice of distance measure
in phase space and are therefore variant under spatial coordinate
transformations also. From the above discussion it is clear that standard
coordinate dependent measures of chaos have to be either modified,
abandoned or augmented in general relativity\cite{rugh}.

This problem has now been solved in a series of papers\cite{dfc,cl,cg}
which introduced and illustrated the effectiveness of fractal
dimensions and symbolic codings as invariant descriptors of chaos in
general relativity. Central to both of these methods is the
concept of a chaotic invariant set of orbits\cite{cit}. It is
interesting to note that the problem of describing chaos in quantum
mechanics was also solved by focusing on periodic orbits\cite{gut}.
In this talk I review these developments and
illustrate the methods by applying them to photon orbits in a binary
black hole system.

\section{Invariant methods}

In order to be precise, I will begin by stating the definitions and
theorems that underpin our approach. I will then illustrate the ideas
pictorially for photon trajectories in the field of two fixed extremal
black holes\cite{cg}.

\noindent {\bf Definition:} A dynamical system will be called chaotic if it
contains a {\em chaotic invariant set} of unstable periodic
orbits. This set is characterised by having a non-zero topological
entropy and non-integer fractal dimensions. Conversely, the system is
integrable if there exist sufficient constants of motion
to restrict all trajectories to smooth tori in phase space.

\noindent {\bf Definition:} A future invariant set forms the fractal
boundary between possible outcomes of a scattering system. The set
consists of trajectories that are future asymptotic to unstable
periodic orbits.

\noindent \underline{{\bf Theorem:}} (see eg. Barnsley\cite{barn})\\
Fractal dimensions are invariant under coordinate transformations
that are one to one, onto and Lipschitz ($C^{1-}$), {\it ie.} bounded
deformations with no rips, tears, folds or infinite stretching. Thus,
fractal dimensions are diffeomorphism invariant.

\begin{figure}[h]
\vspace*{2.2in}
\includegraphics{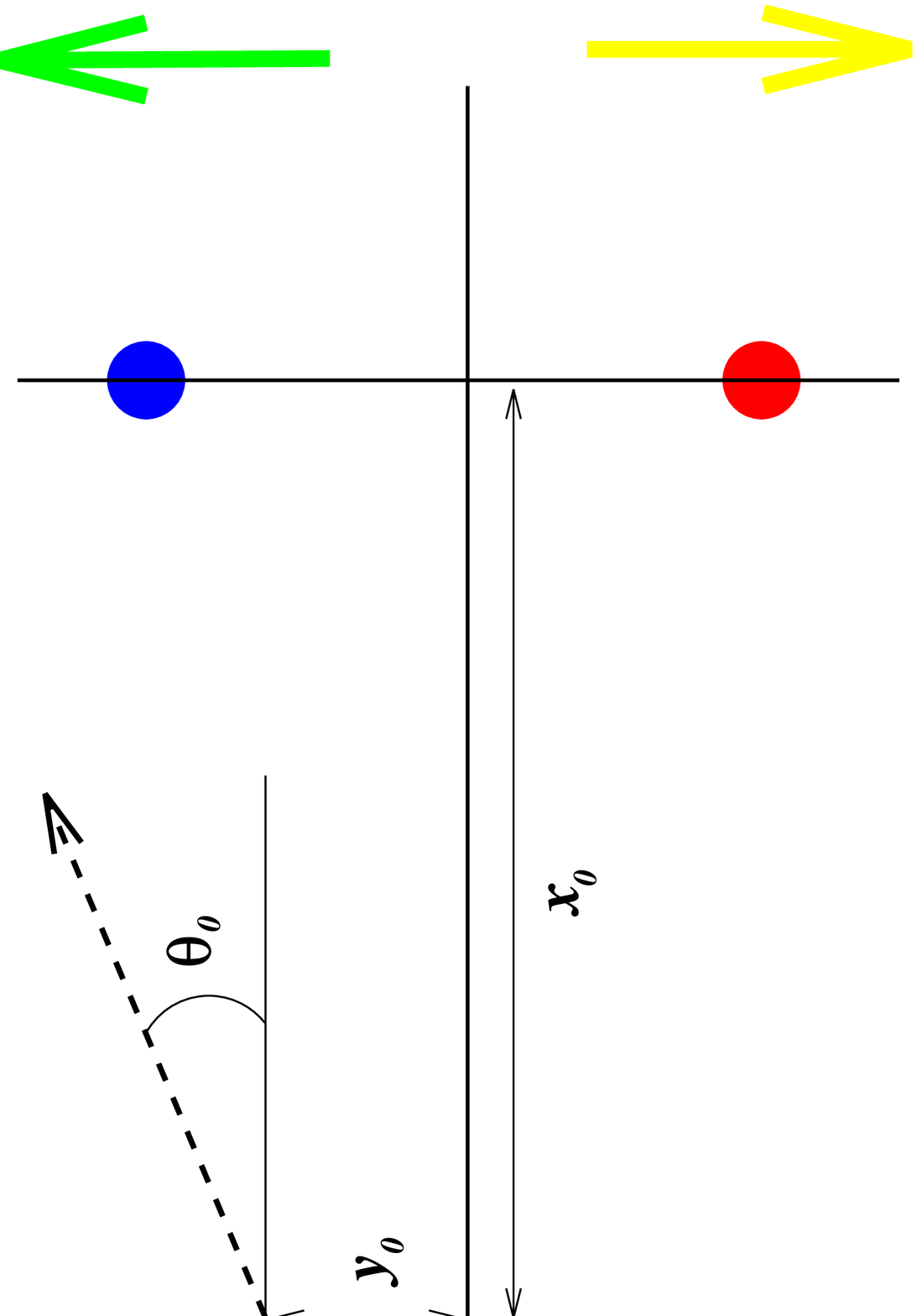}
\includegraphics{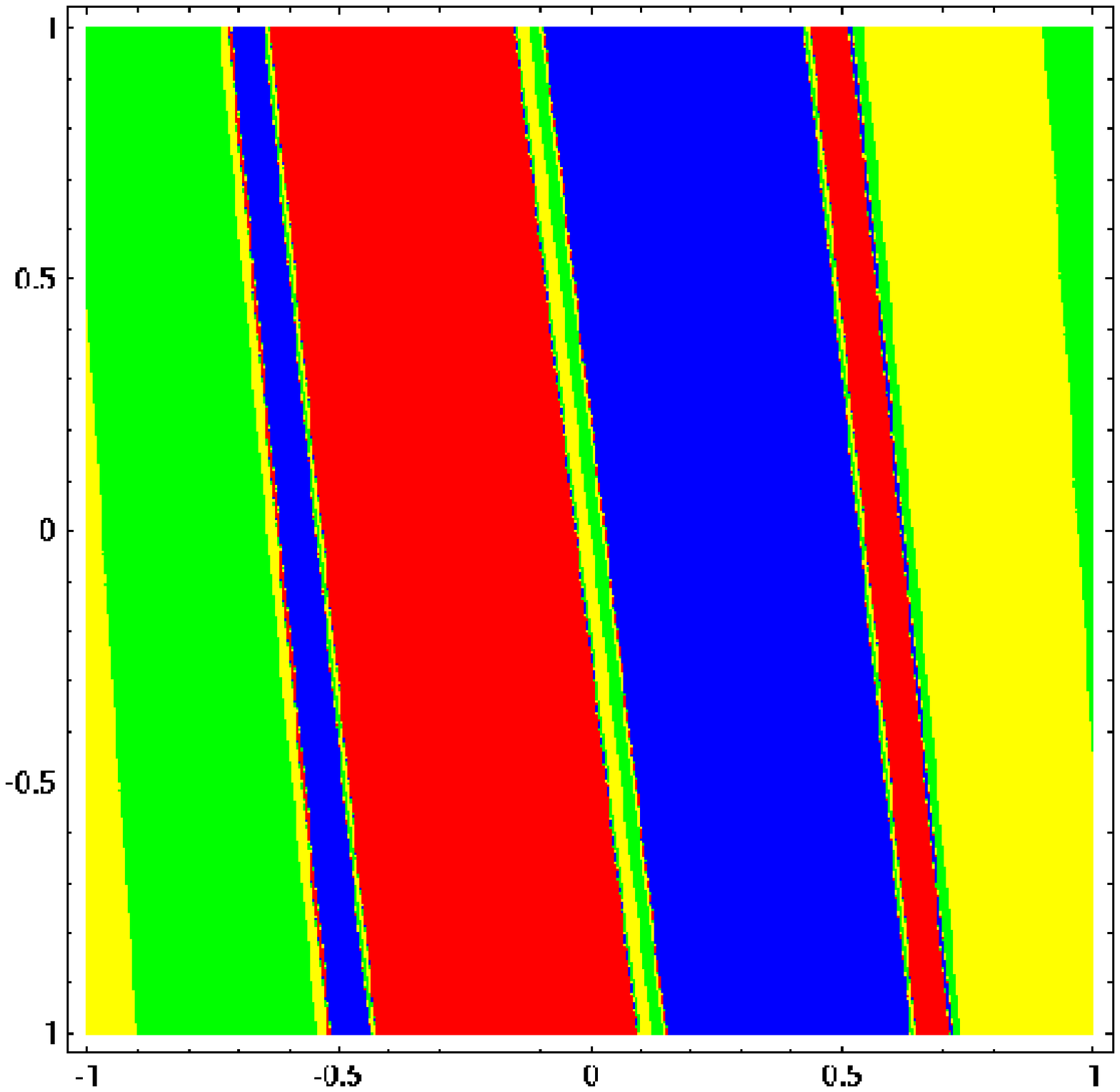}
\caption{(i) Outcomes for an incident photon (ii)
Outcome basins with fractal boundaries.}
\end{figure}
\begin{picture}(0,0)
\put(-10,190){(i)}
\put(170,190){(ii)}
\put(175,123){${ y_0}$}
\put(265,36){${ \theta_0}$}
\end{picture}
\vspace*{-0.1in}

The procedure we adopt is very simple. First we identify the possible
outcomes for the dynamics. Figure (1.i) shows the four outcomes that can
occur when a photon is fired into the field of two fixed, extremal
black holes. The photon is either captured by one of the black holes,
or it escapes and scatters up or down. The outcomes are each assigned a
different colour and the initial conditions are colour
coded according to their outcome. The result of applying this procedure
to a $840\times840$ grid of initial conditions is shown in
Fig.~(1.ii). The boundaries between the various outcomes form a
fractal future invariant set with capacity dimension\cite{cg}
$D_0=1.36\pm 0.02$. Since the set has a non-integer capacity
dimension, we can conclude in a coordinate invariant way that
photon dynamics in this spacetime is chaotic.

An alternative way of showing that a system is chaotic is to find a
symbolic coding for all the periodic orbits and then show that the coding
is complex. The philosophy behind this approach being that chaos is a
global phenomenon, so it can be studied by reducing the detailed local
information required to specify a trajectory down to a discrete set of
symbols\cite{morse}. This is accomplished by dividing phase space into
a finite number of partitions and labelling each partition with a
symbol. Trajectories are then represented by strings of these symbols.

\begin{figure}[h]
\vspace*{1.8in}
\includegraphics{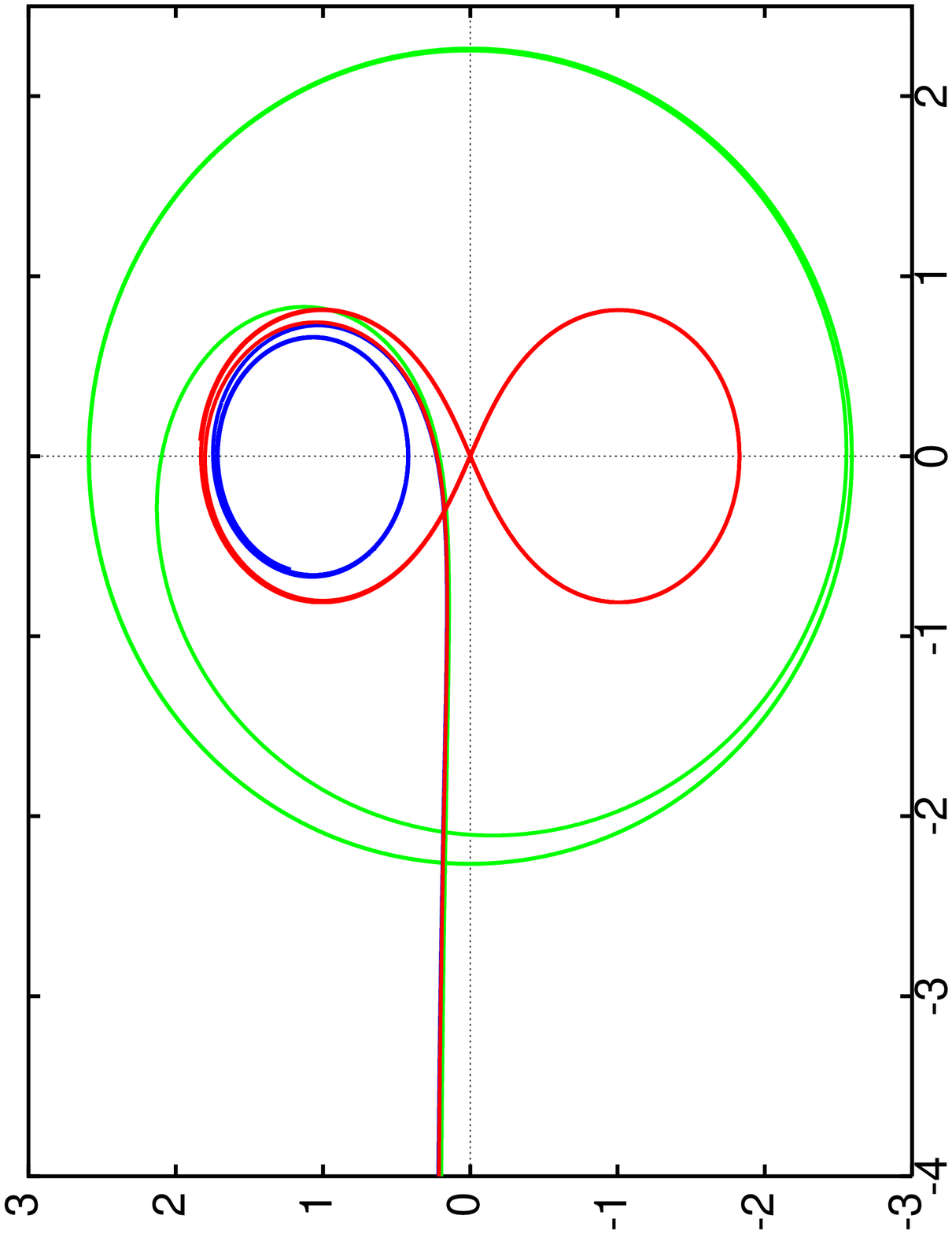}
\includegraphics{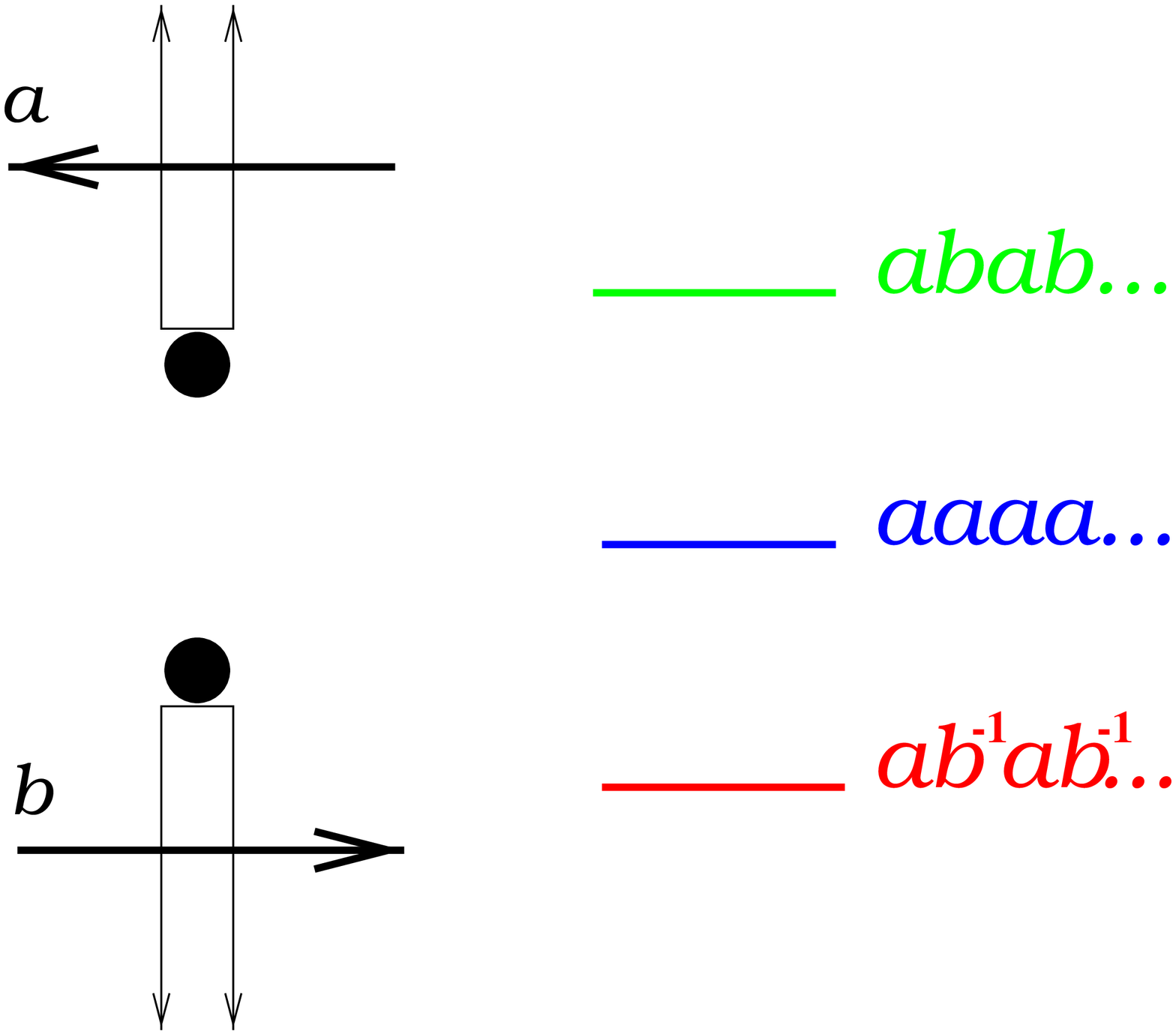}
\caption{(i) Orbits belonging to the future invariant set. (ii) Their
symbolic coding.}
\end{figure}
\begin{picture}(0,0)
\put(5,150){(i)}
\put(260,150){(ii)}
\end{picture}
\vspace*{-0.2in}

Figure (2.i) shows three photon trajectories that take up unstable
orbits around the binary black hole system. An efficient symbolic
coding for the periodic orbits is shown in Fig.(2.ii), along with the
codings for the three simplest orbits shown in Fig.(2.i). The number
of orbits with period $k$ follows from the recurrence relation
$N(k+1)=3 N(k)$. A symbolic coding is said to be grammatically complex
if the number of orbits, ${\cal N}(L)$, with period less than or equal
to $L$ grows exponentially. The growth in the number of orbits is
measured by the topological entropy: $H_T = \lim_{L\rightarrow \infty}
 L^{-1} \ln {\cal N}(L) \, .$
For photon orbits in the field of two extremal black holes we find
$H_T=\ln 3$, so the dynamics is chaotic.

\section*{Acknowledgments}
The work reported in this talk is the result of some very enjoyable
collaborations with Carl Dettmann, Norm Frankel, Gary Gibbons and
Janna Levin.

\end{document}